\documentclass[prl, twocolumn, preprintnumbers, amsmath, amssymb, superscriptaddress]{revtex4}
\usepackage{graphicx}
\usepackage{dcolumn}
\usepackage{bm}
\usepackage{subfigure}
\usepackage{multirow}
\usepackage{amsmath}
\usepackage{color,soul}
\usepackage[font=footnotesize, justification=raggedright,singlelinecheck=false]{caption}
\usepackage[dvipsnames]{xcolor}
\usepackage{mathrsfs}
\usepackage{mathtools}
\usepackage{relsize}

\usepackage[all,cmtip]{xy}
\usepackage[normalem]{ulem}

\begin{document}
\title{Fluidization and anomalous density fluctuations in 2D Voronoi cell tissues with pulsating activity}

\author{Zhu-Qin Li}
\affiliation{National Laboratory of Solid State Microstructures and School of Physics, Collaborative Innovation Center of Advanced Microstructures,  Nanjing University,  Nanjing 210093, People's Republic of China}

\author{Qun-Li Lei}
\email{lql@nju.edu.cn}
\affiliation{National Laboratory of Solid State Microstructures and School of Physics, Collaborative Innovation Center of Advanced Microstructures,  Nanjing University,  Nanjing 210093, People's Republic of China}
\affiliation{Jiangsu Physical Science Research Center, Nanjing 210093, People's Republic of China}
\affiliation{Hefei National Laboratory, Hefei 230088, People's Republic of China}

\author{Yu-qiang Ma}
\email{myqiang@nju.edu.cn}
\affiliation{National Laboratory of Solid State Microstructures and School of Physics, Collaborative Innovation Center of Advanced Microstructures,  Nanjing University,  Nanjing 210093, People's Republic of China}
\affiliation{Jiangsu Physical Science Research Center, Nanjing 210093, People's Republic of China}
\affiliation{Hefei National Laboratory, Hefei 230088, People's Republic of China}

\begin{abstract}
Cells not only can be motile by crawling but are also capable of non-motility active motions like periodic contraction or pulsation. In this work, based on a Voronoi cell model, we show how this non-motility activity affects the structure, dynamic and density fluctuations of cellular monolayers. Our model shows that random cell pulsation fluidizes solid epithelial tissues into a hyperuniform fluid state, while pulsation synchronization inhibits the fluidity and causes a reverse solidification. Our results indicate this solidification is a BKT-type transition, characterized by strong density/dynamic heterogeneity arising from the annihilation of topological defects in the pulsating phase space. The magnitude and length scale of density heterogeneity diverge with the pulsating period, resulting in an opposite giant density fluctuation or anti-hyperuniformity. We propose a fluctuating hydrodynamic theory that can unify the two opposite anomalous fluctuation phenomena. Our findings can help to understand recent experimental observations in MDCK monolayer.
\end{abstract}
\maketitle

\section{Introduction}
Collective motion of cells in epithelial tissues underlies many vital life processes, like embryonic development, wound healing and cancer invasion~\cite{friedl2009collective,hakim2017collective,alert2020physical}. 
{These processes  are usually associated with the fluidization of epithelial tissues~\cite{park2015unjamming, bi2015density, bi2016motility, mongera2018fluid} or flocking of cells~\cite{lin2018dynamic,giavazzi2018flocking}, etc.~\cite{petrolli2019confinement, peyret2019sustained, yu2021spatiotemporal}. Previous studies focused on the role of crawling-induced motility of cells in these processes~\cite{bi2016motility,barton2017active,lin2018dynamic,giavazzi2018flocking,petrolli2019confinement,pasupalak2020hexatic,cai2022numerical,amiri2023random}.}
However, non-motility activities, e.g., active cell deformation or periodic pulsation~\cite{tjhung2017discontinuous,zhang2023pulsating, manacorda2023pulsating,manning2023essay,liu2024collective,pineros2024biased}, are also generic features of cell tissues, which generate active reciprocal force or stress between cells. For example, pulsating contraction induced by actin and myosin dynamics~\cite{martin2009pulsed,dierkes2014spontaneous,lin2017activation,curran2017myosin,krajnc2021active,perez2024excitable} are closely related to morphogenesis, pattern formations and contraction waves in epithelial tissues~\cite{solon2009pulsed, banerjee2015propagating, armon2018ultrafast,krajnc2020solid,hino2020erk,boocock2021theory,devany2021cell,boocock2023interplay,yamamoto2022non,duclut2022active,staddon2022pulsatile,thiagarajan2022pulsations,perez2023tension,yin2024emergence,liu2024emergence}. Meanwhile, pulsating activity can be caused by the exchange of fluid between cells~\cite{zehnder2015cell,zehnder2015multicellular}, or active nuclear movements~\cite{bocanegra2023cell}. During these active pulsating processes, synchronization has been observed~\cite{zehnder2015cell,boocock2023interplay}.  How pulsating activities and their synchronization affect the fluidity of epithelial tissues remains poorly understood. Moreover, epithelial tissues show \emph{giant} density fluctuations driven by collective cell migration or cell pulsation~\cite{zehnder2015cell,giavazzi2017giant}, while in other cellular systems, suppressed density fluctuations or \emph{hyperuniform} states were reported ~\cite{jiao2014avian,chen2016structural,li2018biological, zheng2020hyperuniformity,liu2024universal}.  Thus, it is also intriguing to reveal the connection between pulsating activities and anomalous density fluctuations in epithelial tissues.

In this work, we theoretically study the effects of pulsating activity on the structure, dynamic, and density fluctuations of epithelial monolayers. Based on a Voronoi cell model, we find that random cell pulsation fluidizes epithelial tissues, resulting in a hyperuniform fluid state of {class \uppercase\expandafter{\romannumeral 1}} hyperuniformity~\cite{torquato2003local,torquato2018hyperuniform,lei2019nonequilibrium, lei2019hydrodynamics}. When cell pulsations are synchronized, the fluid state is reversely solidified.  Our analysis indicates this synchronization-induced solidification is a Berezinskii-Kosterlitz-Thouless (BKT) type transition, where density and dynamic heterogeneity emerge at the transition point due to the birth of topological defects in the pulsating phase space. Moreover, we find both the magnitude and length scale of the density heterogeneity diverge with the pulsating period, which results in an opposite giant density fluctuation~\cite{ramaswamy2003active,shi2013topological} or anti-hyperuniformity~\cite{torquato2018hyperuniform,newby2024structural}. We generalize the underlying mechanism of the apparently opposite fluctuation phenomena into a fluctuating hydrodynamic theory. We also demonstrate that above phenomena are insensitive to the form of pulsation, e.g., cell volume or tension pulsation. Our work provides a minimal model to understand recently observed giant density fluctuations and associated topological defects in MDCK monolayer~\cite{zehnder2015cell,boocock2021theory, boocock2023interplay}.

\begin{figure}[!htb] 
\centering
\begin{tabular}{c}
	\resizebox{80mm}{!}{\includegraphics[trim=0.0in 0.0in 0.0in 0.0in]{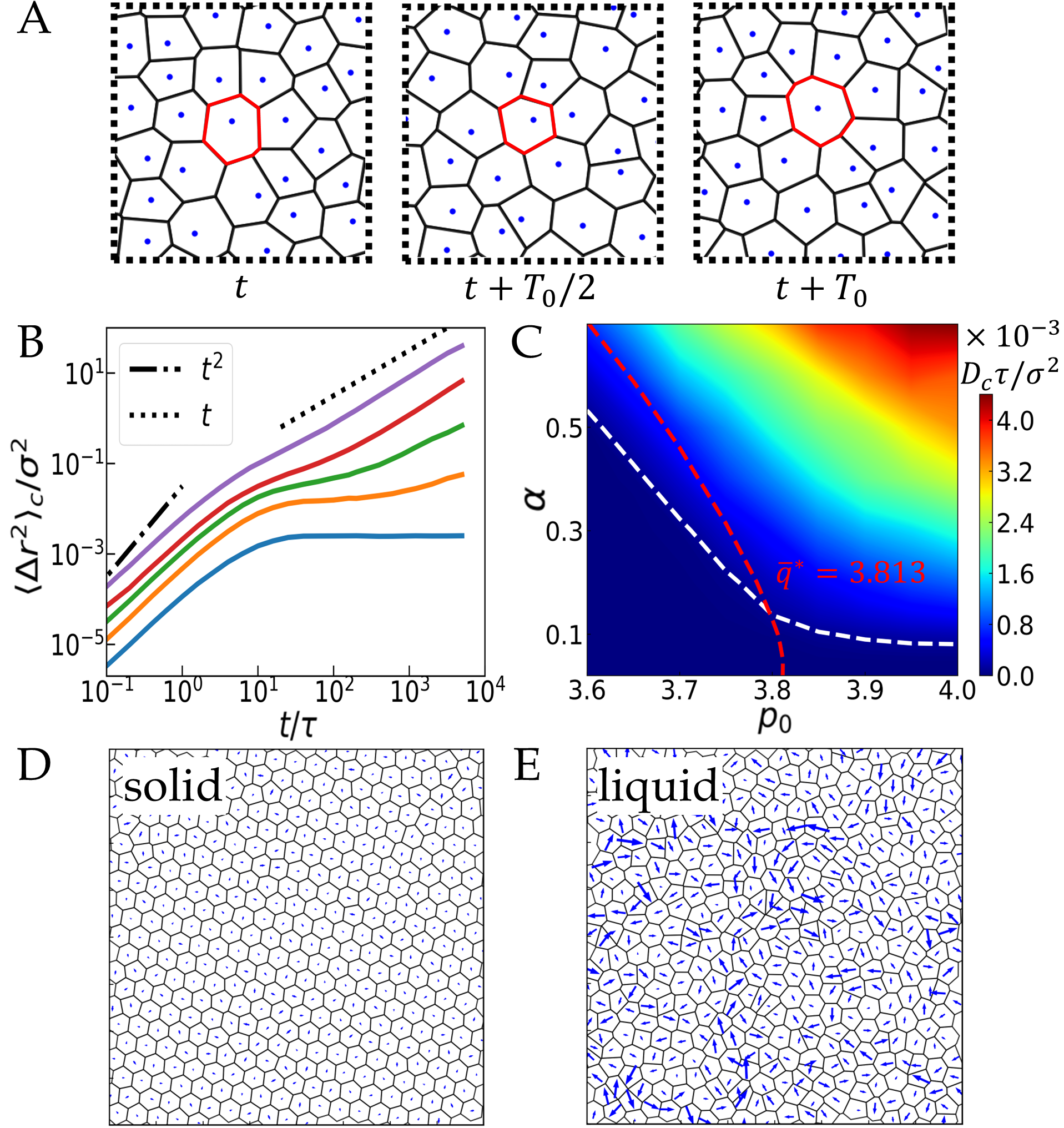} }
\end{tabular}
\caption{(\emph{A}) Schematics of a cell configuration (denoted in red) over a cell pulsating cycle $T_0$. (\emph{B}) ${\rm MSD}_c$ under fixed $p_0=3.7$ and different pulsating strength $\alpha=[0.1,\ 0.2,\ 0.3,\ 0.4,\ 0.7]$ (bottom to top).  (\emph{C}) Phase diagram of diffusion coefficient $D_{\rm{c}}$. The white dashed line represents  $D_{\rm{c}}^*=10^{-4}\sigma^2/\tau$.  (\emph{D-E}) Typical solid ($p_0=3.7,\ \alpha=0.2$) and fluid ($p_0=3.7,\ \alpha=0.6$) configurations. Blue arrows show the cell displacements after one period.  All data are for random pulsation systems ($J=0$). }
\label{fluidization}
\end{figure}

\section*{Results}
\subsection*{Model}

We consider a 2D confluent cell monolayer composed of $N$ pulsating cells whose area and shape can change with time (Fig.~\ref{fluidization}A). Cells are modelled as polygons determined by the Voronoi tessellation of the space~\cite{bi2016motility}. We assume the pulsation is strong enough that the effect of thermal noise can be neglected in the overdamped cell dynamics. In our model, each cell $i$ has a normalized preferred area $a^i_0$ which does stochastic oscillation, i.e.,
\begin{eqnarray}
a^i_0 = 1 + \alpha \sin\varphi_i(t),
\label{preferred area}
\end{eqnarray}
where $\alpha$ is the pulsating strength and $\varphi_i(t)$ is the pulsating phase.  The normalized preferred perimeter is $p^i_0=p_0$ with $p_0$ the so-called \textit{target shape factor}~\cite{bi2015density,bi2016motility} characterizing the time-averaged preferred shape of cells. The preferred area and perimeter of cells can not be all satisfied in a confluent cell tissue, which gives rise to an effective energy that determines the dynamics of cells (see Materials and Methods for details). Moreover, to account for collective cell pulsation, we assume $\varphi_i(t)$ obeys the stochastic Kuramoto-like dynamics~\cite{kuramoto1975international,zheng1998phase,acebron2005kuramoto,rouzaire2021defect}, i.e.,
\begin{eqnarray}
\frac{d\varphi_i}{dt} = \omega_0 + J\sum_{j \in C_i}\sin(\varphi_j-\varphi_i)+\sqrt{2T_\varphi}\eta_i(t), \label{Kuramoto model}
\end{eqnarray}
where frequency $\omega_0$ and the corresponding period $T_0=2\pi/\omega_0$ are supposed to be associated with certain intrinsic cell cycles~\cite{zehnder2015cell,boocock2023interplay}. The second term represents the coupling between adjacent cells with $J$ the coupling strength. $T_\varphi$ is the noise strength in the phase space and $\eta_i(t)$ is the Gaussian white noise. This simplified model is inspired by more sophisticated biological models while retaining the essential physics. It should be noted that Eq.~(\ref{Kuramoto model}) is decoupled from the cellular interaction energy  (Eq.~(\ref{energy}) in Materials and Methods). Adding such a coupling essentially enables the mechanochemical feedback which can give rise to more complicated cellular dynamics~\cite{boocock2023interplay,zhang2023pulsating,pineros2024biased,banerjee2024hydrodynamics}. In this article, we define the unit length and unit time as $\sigma$ and $\tau$, respectively. Detail simulation information can be found in the Materials and Methods.

\begin{figure*}[!htb] 
\centering
\begin{tabular}{c}
	\resizebox{165mm}{!}{\includegraphics[trim=0.0in 0.0in 0.0in 0.0in]{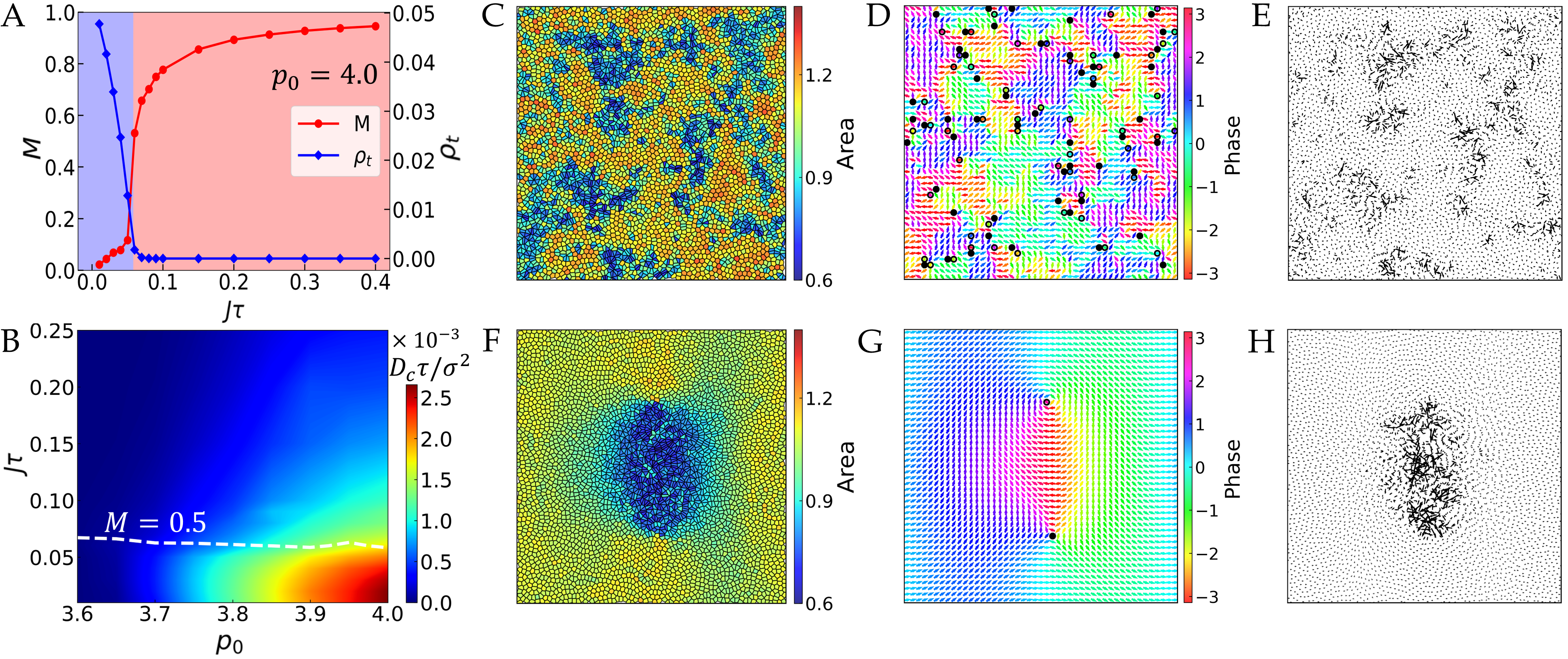} }
\end{tabular}
\caption{(\emph{A}) Phase polarization $M$ (red line) and topological defect density $\rho_t$ (blue line) as a function of $J$ at fixed $p_0 = 4.0$,  with transition point $J^*\simeq 0.06\tau^{-1}$. (\emph{B}) Phase diagram of $D_c$ at $\alpha=0.4$. The white dashed line represents states with $M=0.5$. (\emph{C}) Cell area distribution for the system at $J=0.05\tau^{-1}$ and $p_0=4.0$. (\emph{D}) The pulsating phase field for the same configuration in (\emph{C}) with the color arrows indicates the phase. The solid circle and open circle represent the $+1$ and $-1$ topological defects, respectively.  (\emph{E}) Instantaneous velocity field for the configuration in (\emph{C}). (\emph{F-H}) The same figures as that in (\emph{C-E}) for a pair of topological defects at $J=0.01 \tau^{-1}$, $p_0=4.0$ and $T_\varphi=0$. For all simulations, $N=4096$.}
\label{synchronization}
\end{figure*} 

\subsection*{Cell-pulsation induced fluidization}
We first study the case of zero pulsation coupling ($J=0$). To characterize the dynamical states of cell tissues, we calculate the cage relative mean square displacement of cells, ${ {\rm MSD}_{c} }(t)$, which excludes the influence of the long-wavelength fluctuations in 2D (Materials and Methods).  As shown in Fig.~\ref{fluidization}B, the plateau of ${ {\rm MSD}_{c} }$ gradually disappears and exhibits diffusion scaling with increasing $\alpha$. This indicates a pulsation-induced fluidization analogous to the one induced by motility~\cite{bi2016motility}. Similar behaviors have been reported in the actively deforming spherical particle system~\cite{tjhung2017discontinuous}. Microscopically, the fluidization arises from unbalanced intercellular forces induced by random pulsation, which results in frequent T1 transitions (Movie S1).  In Fig.~\ref{fluidization}C, we plot the diffusion coefficient $D_{\rm{c}}$ of systems in dimensions of $\alpha$ and $p_0$, where the solid and liquid phases can be roughly separated by a threshold  $D_{\rm{c}}^* = 10^{-4}\sigma^2/\tau$~\cite{bi2016motility}. We also confirm that the solid phase in our model has the crystal order (\emph{SI Appendix}, Fig.~S1B and Movie S2).  It would be interesting to further study the nature of this transition and the glass dynamics in detail~\cite{li2024relaxation,ansell2024tunable}.  From Fig.~\ref{fluidization}C, one can see that systems with large $p_0$ are more easily fluidized as long as $\alpha \gtrsim 0.1$. This accords with the energy barrier scenario of T1 transition~\cite{bi2015density}. Nevertheless, by accurately calculating the average shape factor $\bar{q} = \sum_i(p_i/\sqrt{a_i})/N$, we find the line of zero-energy-barrier, i.e., $\bar{q}^*=3.813$~\cite{bi2015density,bi2016motility}, does not  follow  the trend of  line   $D_{\rm{c}}^*$, especially for small $\alpha$. {This results in a phase regime  with $p_0>3.8$ and $\alpha<0.1$, in which cells are highly deformed (large $\bar{q}$) but their movements are arrested (small $D_c$).} Similar states have also been observed experimentally in MDCK monolayer with low fluctuating cell junction tension~\cite{devany2021cell}.

\subsection*{Synchronization-induced solidification}
We further consider the intercellular coupling during pulsation. We find that increasing coupling strength $J$ leads to a synchronization transition at $J^* \simeq 0.06 \tau^{-1}$, {above which cells pulsation are synchronized. The  synchronization degree} can be characterized by the phase polarity $M = \langle |\sum_i (\cos \varphi_i, \sin \varphi_i )| \rangle/N$ (Fig.~\ref{synchronization}A), {which ranges from [0, 1]}.  We find that this synchronization transition is insensitive to the choice of $\alpha$ and $p_0$ (\emph{SI Appendix}, Fig.~S2A, C).  To explore how the synchronization transition in the pulsating phase space affects the cell dynamics in real space,  we further plot $D_{\rm{c}}$ in dimensions of $J$ and $p_0$ at fixed $\alpha=0.4$ in Fig.~\ref{synchronization}B. We find that for the liquid phase, $D_{\rm{c}}$ drops quickly as $J$ increases above $J^*$, suggesting a synchronization-induced solidification. Compared with the unsynchronized states, the cell sizes in synchronized states are more uniformly distributed (\emph{SI Appendix}, Fig.~S2D). Thus, cells experience a more balanced intercellular force during collective pulsation. This suppresses the T1 transition and leads to the solidification.

\begin{figure*}[!htb]
\centering
\begin{tabular}{c}
	\resizebox{175mm}{!}{\includegraphics[trim=0.20in 0.0in 0.0in 0.0in]{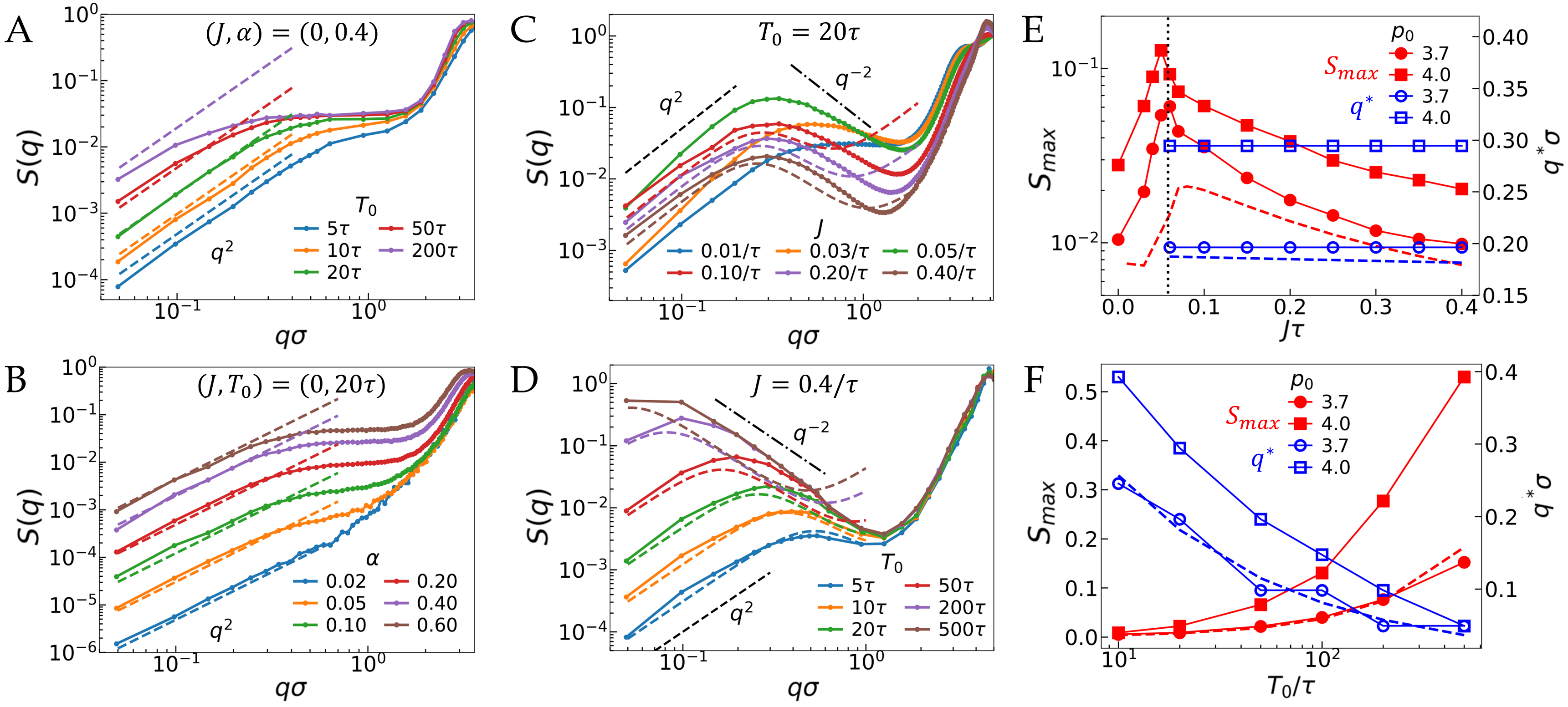} }
\end{tabular}
\caption{(\emph{A-D}) Simulation data (solid symbols) and theoretical predictions (dashed lines) of $S(q)$ for systems with: (\emph{A}) different $T_0$ at fixed $(p_0,\ \alpha) = (4.0,\ 0.4)$ and (\emph{B}) different $\alpha$ at fixed $(p_0,\ T_0)=(4.0,\ 20\tau)$, under random pulsation  ($J=0$); (\emph{C}) different $J$ at fixed  $(p_0,\ \alpha,\ T_0)=(4.0,\ 0.4,\ 20\tau)$; (\emph{D}) different $T_0$ at fixed $(J,\ p_0,\ \alpha) = (0.4\tau^{-1},\ 4.0,\ 0.4)$. In (\emph{C}) and (\emph{D}), a correction $\chi=0.25$ is used; (\emph{E-F}) Simulation data of $S_{max}$ (red solid symbols) and $q^*$ (blue hollow symbols)  as well as the corresponding theoretical predictions (dashed lines) for systems with: (\emph{E}) different $J$ at fixed $(\alpha,\ T_0)=(0.4,\ 20\tau)$; (\emph{F}) different $T_0$ at fixed $(J,\ \alpha) = (0.4\tau^{-1},\ 0.4)$. Note that no $\chi$ correction is used in the theoretical predictions in (\emph{E}) and (\emph{F}). Here,  $N=16384$ for all simulations. $\gamma=1.1$ and simulation data of $M$ are used for all theoretical predictions.}
\label{Anomalous-DF}
\end{figure*}

Interestingly, near the synchronization transition point, we find the coexistence of high-density contracted regions and low-density expanded regions that periodically switch their states (Fig.~\ref{synchronization}C and Movie S3). This density heterogeneity is strongest at $J^*$, as evidenced by the height of the first peak in structure factor $S(q)$ (Fig.~\ref{Anomalous-DF}C). In Fig.~\ref{synchronization}E, we also show that cell motility is negligible like a solid in the expanded regions, while much higher akin to a liquid in the contracted regions.  We find that this dynamic heterogeneity is a result of heterogeneous distribution of local shape factor $\bar{q}_i (t) = (p_i(t)/\sqrt{a_i(t)})$ which is large (small) in the contracted (expanded) regions. A larger (smaller) shape factor corresponds to a more anisotropic (isotropic) cell shape and stronger (weaker) imbalanced intercellular forces which favour the liquid (solid) phase~\cite{bi2015density,bi2016motility}. We note that similar density heterogeneities and local synchronized pulsation patterns were observed in MDCK monolayers~\cite{zehnder2015cell, hino2020erk, boocock2021theory}. 

It is known that the (de-)synchronization transition described by lattice Kuramoto model with monodispersed frequency belongs to BKT universality, accompanied by the creation and annihilation of ($\pm 1$) topological defects~\cite{rouzaire2021defect}. We confirm this scene by calculating the topological defect density $\rho_t=Q/N$ in the pulsating phase space (Fig.~\ref{synchronization}A). Furthermore, we plot the pulsating phase field of the cell monolayer near the critical point in Fig.~\ref{synchronization}D. We identify  $\pm 1$  topological defects and find they are mostly located on the boundaries between contracted and expanded regions. This kind of topological defects has recently been identified in MDCK monolayer in experiments~\cite{boocock2023interplay}. To further disclose the relationship between the topological defects and density/dynamic heterogeneities, we create a single pair of  $\pm 1$ topological defects in the pulsating phase space and observe its evolution (Fig.~\ref{synchronization}F-H). We find that the topological defects create a synchronized cell droplet in between, which contracts (fluidizes) or expands (solidifies) alternatively.  It also creates a local pulsating wave around the defect pair (Movie S4). Thus, the synchronization-induced solidification of the monolayers is essentially a BKT-type transition, resulting from the annihilation of topological defects in the pulsating phase space. 

\subsection*{Anomalous density fluctuations}
\begin{figure}[!htb]
\centering
\begin{tabular}{c}
	\resizebox{75mm}{!}{\includegraphics[trim=0.0in 0.0in 0.0in 0.0in]{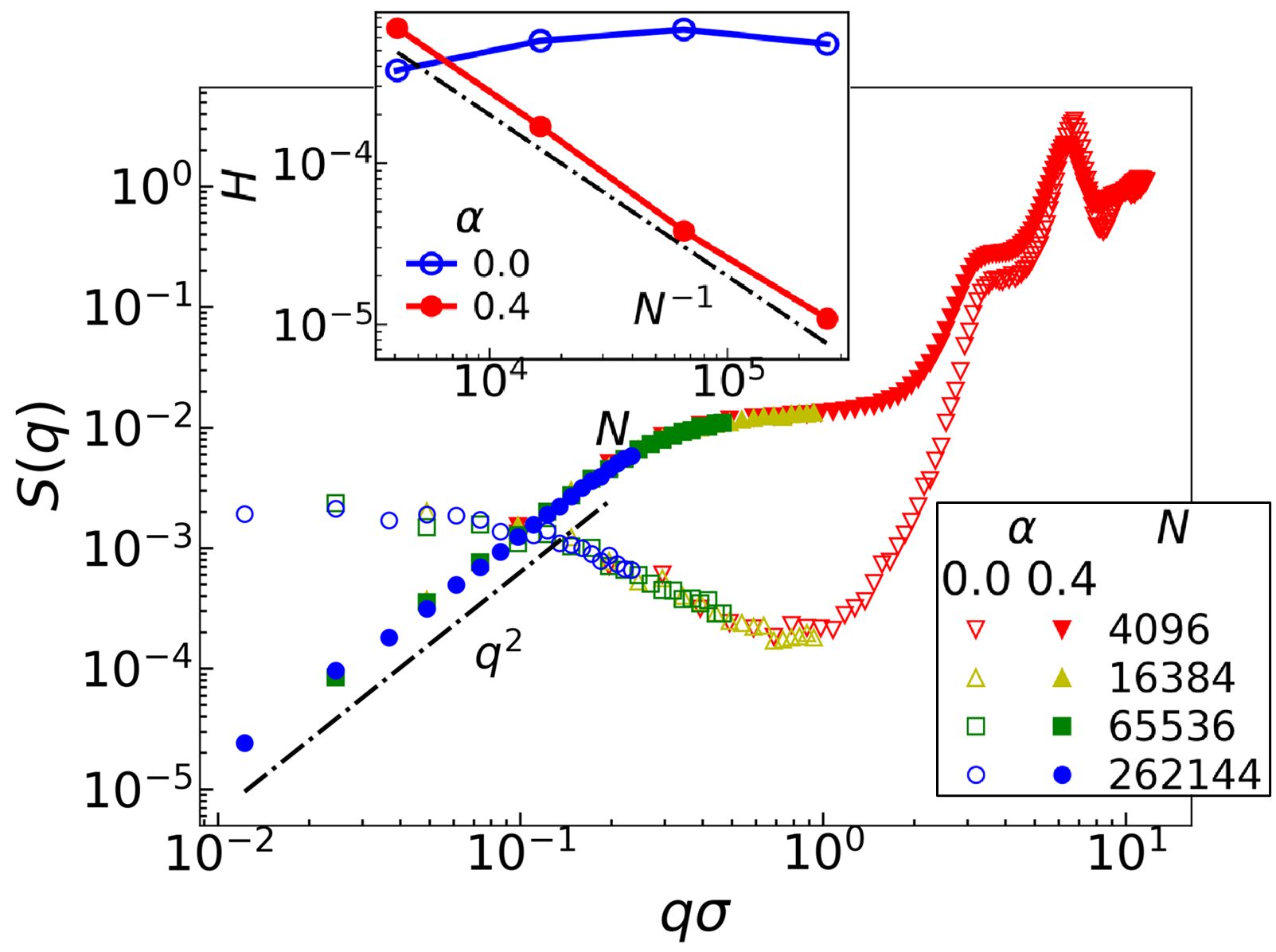} }
\end{tabular}
\caption{Finite size effect analysis of static structure factors for the system with random pulsation ($\alpha=0.4$, solid symbols) and without pulsation ($\alpha=0$, open symbols). Inset: The effective hyperuniformity index $H=S(q \rightarrow 0)/S(q_{peak})$ as a function of $N$ for the two cases. Other parameters are $(p_0,\ T_0)= (3.8,\ 20\tau)$.}
\label{Compare}
\end{figure}

Next, we focus on the density fluctuations of the pulsation system, which can be characterized by the static structure factor $S(q)$. As shown in Fig.~\ref{Compare},  $S(q)$ for random pulsating tissue ($J=0$) in liquid state ($\alpha=0.4$) shows a scaling $S(q \rightarrow 0) \sim q^2$, which is a hallmark of class \uppercase\expandafter{\romannumeral 1} hyperuniform state with vanishing density fluctuations at infinite wavelength~\cite{torquato2003local,torquato2018hyperuniform}.  Similar hyperuniform fluid states were also found in circle swimmers and active spinners systems~\cite{lei2019nonequilibrium, lei2019hydrodynamics, huang2021circular}.  In order to study the effect of pulsating period $T_0$ on hyperuniformity,  we plot the structure factor $S(q)$ for random pulsation systems under different $T_0$ and fixed $(p_0,\ \alpha)=(4.0,\ 0.4)$ in Fig.~\ref{Anomalous-DF}A. With decreasing $T_0$, we find that the $q^2$ scaling remains unchanged, while the degree of hyperuniformity is enhanced. Decreasing $\alpha$ has a similar effect (Fig.~\ref{Anomalous-DF}B), but the relaxation dynamic would also be slower. Thus, hyperuniformity is the strongest when cells undergo fast and weak pulsation in the liquid state.

In addition to the pulsating period $T_0$ and strength $\alpha$, the synchronization of pulsation also affects the density fluctuations of the system. With increasing $J$ from zero to $J^*$ for system under $(p_0,\ \alpha)=(4.0,\ 0.4)$, we find that a new peak in $S(q)$ gradually develops around $q^* \sim 0.3~\sigma^{-1}$ without modifying the $q^2$ scaling at small $q$ regime (Fig.~\ref{Anomalous-DF}C). This indicates the coexistence of local large density fluctuations with global hyperuniformity, a scenario similar to circle swimmer systems~\cite{lei2019nonequilibrium}. Interestingly, with further increasing $J$, the height of the peak denoted as $S_{max}$ gradually decreases. This is best shown in Fig.~\ref{Anomalous-DF}E, where we plot $S_{max}$ as a function of $J$ for two different $p_0$.  We find $S_{max}$ exhibits diverging behavior when $J \rightarrow J^*$, which is more pronounced for large $p_0$. In Fig.~\ref{Anomalous-DF}D, we further show $S(q)$ for various $T_0$ in synchronized states with $J=0.4\tau^{-1}$. With increasing $T_0$, $S_{max}$ increases and the location of the peak $q^*$ shifts to the small $q$. This indicates that both the magnitude and length scale of the density heterogeneity diverge when $T_0 \rightarrow \infty$, resulting in giant density fluctuations or anti-hyperuniformity following the reverse $q^{-2}$ scaling. This mechanism is quite robust, independent of whether the system is in the solid or liquid phase (\emph{SI Appendix}, Fig.~S3). It should be mentioned that giant density fluctuation was reported in MDCK monolayer experimentally~\cite{zehnder2015cell}. Our model is the minimal model that can qualitatively reproduce this phenomenon (Movie S5).

It should be emphasized that the hyperuniform fluid state induced by cell pulsation is distinct from previously reported static effective hyperuniform state  (corresponding to $\alpha=0$ in our model)~\cite{li2018biological, zheng2020hyperuniformity}. As a comparison, in Fig.~\ref{Compare}, we plot the structure factor for systems with random pulsation ($\alpha=0.4$) and without pulsation ($\alpha=0$) under different system sizes and fixed $(p_0,\ T_0)= (3.8,\ 20\tau)$. One can see that while the pulsating cellular system shows system-size independent $q^2$ hyperuniform scaling,  the non-pulsating cellular system ($\alpha=0$)  only shows a certain degree of uniformity. This is further confirmed by the inset of Fig.~\ref{Compare}, where we compare the effective hyperuniformity index $H=S(q \rightarrow 0)/S(q_{peak})$~\cite{zheng2020hyperuniformity} of two systems. One can see that $H$ of pulsating cell systems vanishes as the system size increases ($H(N) \sim N^{-1}$), which is a characteristic of a strict hyperuniformity, while $H$ remains a constant in the non-pulsating case. In fact, it was suggested that the non-pulsating system can only show strict hyperuniformity in the case of zero perimeter modulus ($K_P=0$) due to the equal-area packing of cells~\cite{gabrielli2008tilings,tang2024tunable}, which is difficult to achieve in real cell tissue. Moreover, in \emph{SI Appendix}, Fig.~S4, we demonstrate that this hyperuniform fluid state is dynamically robust to external perturbation. We also give the response of hyperuniform fluid state to thermal noise in \emph{SI Appendix}, Fig.~S5 which accords with ref.~\cite{lei2019nonequilibrium}. In the next section, we will reveal the hydrodynamic mechanism behind the hyperuniform fluid state, as well as the giant density fluctuation state.

\subsection*{Fluctuating hydrodynamic theory of pulsating cell tissue}
To understand the above anomalous density fluctuations, we construct a hydrodynamic theory based on cell density field  $\rho ({\mathbf r},t)$ and pulsating phase field $\varphi({\mathbf r},t)$(\emph{SI appendix}),
\begin{eqnarray}
\partial_t \rho  &=& \frac{\mu}{\rho_0}\nabla\cdot (\rho \nabla P) + A_{\alpha}  \nabla^2\eta({\mathbf r},t), \label{EODF} \\
\partial_t \varphi &=& \omega_0 + \sqrt{3}\frac{J}{\rho_0}\nabla^2 \varphi+ \sqrt{2T_\varphi}\xi({\mathbf r},t), \label{EOPF}
\end{eqnarray}
where,
\begin{eqnarray}
P(\rho,\varphi, M) &=&  -2K_A\left[\frac{1}{\rho} -  \frac{1}{\rho_0}\left(1+  {\alpha} M \sin\varphi \right) \right] \nonumber \\ 
& & +K_Pp_0^2\left(\frac{\sqrt{\rho}}{\sqrt{\rho_0}}-1\right) \label{Pressue}
\end{eqnarray}
is the local pressure field. For a homogeneous state $\rho=\rho_0$, Eq.~(\ref{Pressue}) can be simplified into $P_0=\frac{2K_A}{\rho_0}\alpha M \sin\varphi$, which reflects that pulsation generates periodic active stress in the cell tissue that becomes zero in the random pulsating case ($M=0$) under coarse-graining. The second term in Eq.~(\ref{EODF}) is the center-of-mass conserved noise terms~\cite{hexner2017noise,lei2019hydrodynamics,lei2019nonequilibrium} caused by the active reciprocal interaction, i.e., cell pulsation. The noise strength is $A_{\alpha}=\gamma  \alpha  (1-M) \sqrt{T_0}$ for small $T_0$. Eq.~(\ref{EOPF}) is the coarse-graining of Eq.~(\ref{Kuramoto model}). $\eta$ and $\xi$ are two uncorrelated Gaussian white noise. It should be mentioned that the mechanochemical feedback mechanism can also be considered in the framework of hydrodynamic theory by adding a term like $-\partial_{\varphi_i}\epsilon$ in Eq.~(\ref{EOPF}), which was shown to lead to dynamic instability patterns, like propagating waves~\cite{zhang2023pulsating,banerjee2024hydrodynamics}. By making a weak perturbation around the synchronous homogeneous state, i.e., $\rho(\vec{r},t)=\rho_0 + \delta \rho(\vec{r},t)$, $\varphi(\vec{r},t)=\omega_0 t + \delta \varphi(\vec{r},t)$, we obtain equations in the Fourier space with the linear approximation,
\begin{eqnarray}
i\omega\delta\rho_{\omega,{\mathbf{q}}} &=& -D' q^2 \delta\rho_{\omega,{\mathbf{q}}} - A_\alpha q^2\eta_{\omega,{\mathbf{q}}} \label{FT-EODF} \nonumber \\
& & -\alpha' q^2(\delta\varphi_{\omega+\omega_0,{\mathbf{q}}}+\delta\varphi_{\omega-\omega_0,{\mathbf{q}}}), \\
i\omega\delta\varphi_{\omega,{\mathbf{q}}} &=& -J' q^2\delta\varphi_{\omega,{\mathbf{q}}} + \sqrt{2T_\varphi} \xi_{\omega,{\mathbf{q}}}, \label{FT-EOPF}
\end{eqnarray}
where $[\delta\rho_{\omega,{\mathbf{q}}},\delta\varphi_{\omega,{\mathbf{q}}}]=\int dt e^{-i\omega t} \int d\vec{r} e^{-i\mathbf{q}\cdot \mathbf{r}}[\delta \rho,\delta \varphi]$. $\alpha' = \frac{\alpha M \mu K_A}{\rho_0}$ is the effective pulsation strength.  $J'=\frac{\sqrt{3}}{\rho_0}J$ and $D'=\frac{\mu}{\rho_0}\left( \frac{2K_A}{\rho_0} + \frac{K_Pp_0^2}{2} \right)$ are the effective coupling strength and diffusion coefficient respectively. From Eq.~(\ref{FT-EODF}-\ref{FT-EOPF}), we can obtain
\begin{eqnarray}
S(q) &=& 2T_\varphi\frac{\alpha'^2(J'+D')}{J'D'}\frac{q^2}{(J'+D')^2q^4+\omega_0^2} + \frac{A_\alpha^2 q^2}{2D'}.~~~\label{Theoretical static structure factors}
\end{eqnarray}
In the case of random pulsation ($M=0$ and $\alpha'=0$),  $S(q)$ exhibits a $q^2$ hyperuniform scaling~\cite{torquato2018hyperuniform}, as a result of the emergent long-range hydrodynamic correlation~\cite{lei2019hydrodynamics}. In Fig.~\ref{Anomalous-DF}A-B, we plot the theoretical prediction as the dashed line, which roughly matches the simulation data under different $\alpha$ and $T_0$ by setting the adjustable parameter $\gamma=1.1$ in $A_{\alpha}$. In the case of synchronized state ($M \simeq 1$ and $A_\alpha \simeq 0$), the theory predicts the emergence of the first peak in $S(q)$ at finite $T_0$ and giant density fluctuations with $q^{-2}$ scaling as $T_0 \rightarrow \infty$. In Fig.~\ref{Anomalous-DF}E-F, we plot $S_{max}$ and $q^*$ of these first peaks as a function of $J$ and $T_0$, respectively. One can see that theoretical lines match the simulation data without additional fitting parameters for the synchronized solid phase ($p_0=3.7$). Especially, the theory predicts the right diffusion scaling relationship $q^{*-2}\simeq (J'+D')T_0$. For the large deformation case ($p_0=4.0$), our current theory overestimates the local pressure contributed by the junction tension of cells. Therefore, a correction $\chi(p_0)<1$ is needed to match the simulation data (Fig.~\ref{Anomalous-DF}C-D and \emph{SI appendix}). {From Eq.~(\ref{Theoretical static structure factors}), one can conclude} that the center-of-mass conserved noise arising from active reciprocal interaction and periodically driving are two important ingredients to form hyperuniform fluid in the pulsation cellular systems~\cite{lei2019nonequilibrium,lei2019hydrodynamics,kuroda2023microscopic,ikeda2023harmonic,ikeda2024continuous,keta2024long}.

\subsection*{Generalization to other pulsation form}
\begin{figure}[!htb] 
\centering
\begin{tabular}{c}
	\resizebox{65mm}{!}{\includegraphics[trim=0.0in 0.0in 0.0in 0.0in]{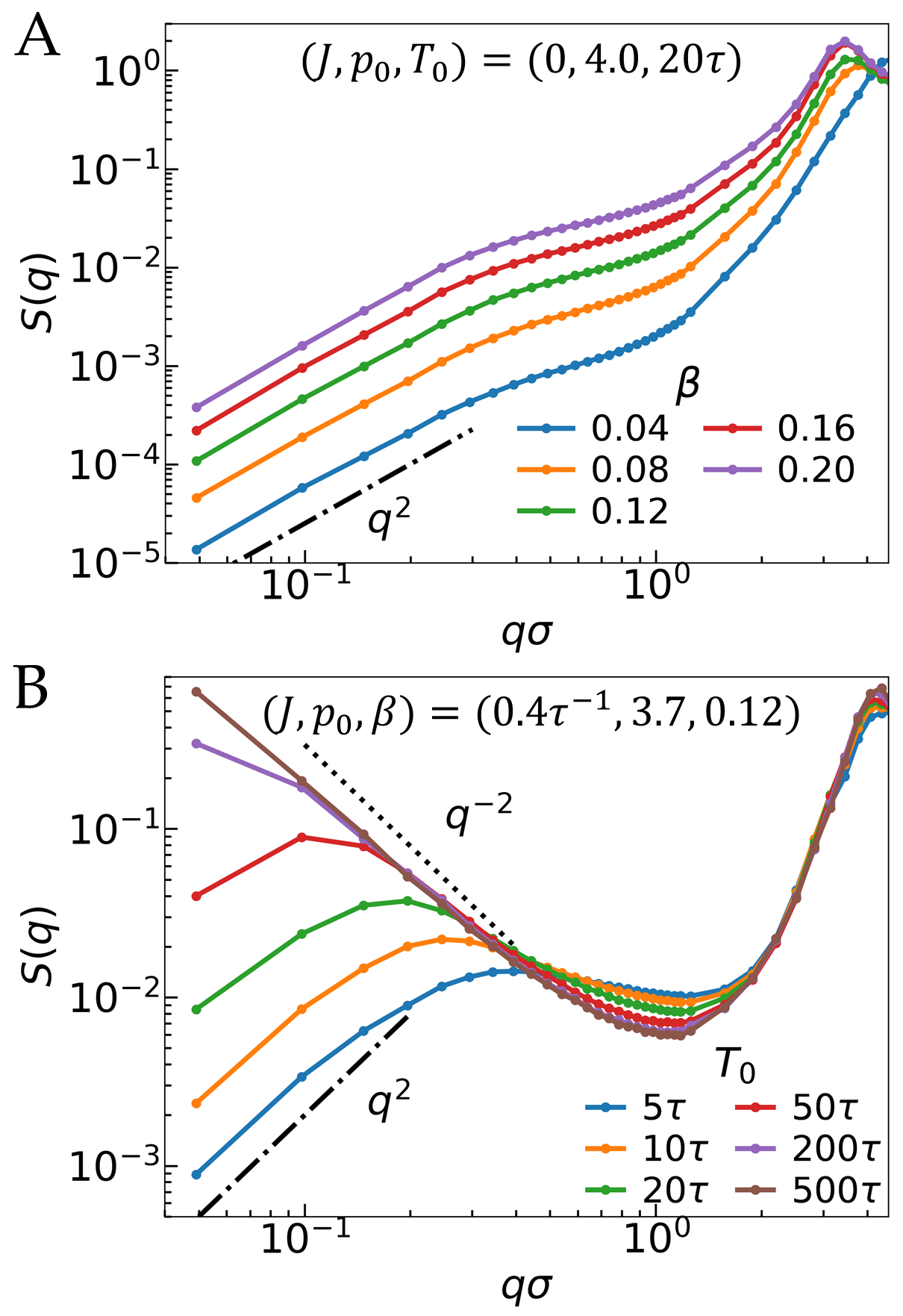} }
\end{tabular}
\caption{Simulation data of $S(q)$ for tension pulsating systems with: \emph{(A)} different pulsation strengths $\beta$ at fixed $(p_0,\ T_0)=(4.0,\ 20\tau)$ for random pulsation case ($J=0$); \emph{(B)} different $T_0$ at fixed $(J,\ p_0,\ \beta) = (0.4\tau^{-1},\ 3.7,\ 0.12)$. Here,  $N=16384$ for all simulations.}
\label{other pulsation}
\end{figure}

In the above, we focus on cell pulsation driven by volume/area oscillation. Biologically, cell pulsation might also come from fluctuating junction tension~\cite{dierkes2014spontaneous,curran2017myosin, krajnc2020solid,krajnc2021active,devany2021cell,yamamoto2022non}. Thus, we extend our studies to the case where $p^i_0$ does stochastic pulsation as 
\begin{eqnarray}
 p^i_0 = p_0[1 + \beta \cos\varphi_i(t)]
 \end{eqnarray}
 while $a^i_0 = 1$. Here, $\beta$ is the pulsating strength. This tension pulsation system exhibits a similar fluidic hyperuniformity under random pulsation (Fig.~\ref{other pulsation}A and \emph{SI Appendix}, Fig.~S6A-B), as well as synchronization-induced dynamical slowdown and giant density fluctuations at large pulsating period (Fig.~\ref{other pulsation}B and \emph{SI Appendix}, Fig.~S6C-D). In addition, topological defects in pulsating phase space still underlie the structure and dynamic heterogeneity of cell tissue (\emph{SI Appendix}, Fig.~S7). Unlike area pulsation, perfectly synchronized tension pulsation is similar to global cyclic shear, which destabilizes the ordered solid state (Movie S6). Finally, the hydrodynamic theory can also be extended to this pulsating scenario and qualitatively explain the two anomalous density fluctuation phenomena (see \emph{SI Appendix} for discussion). All these results demonstrate the robustness of the pulsating activity in determining fluidization and anomalous density fluctuations in cell tissues.

\section*{Discussion and Conclusion}
In this work, we propose an epithelial tissue model with pulsating activity. By using this model, we show that random cell pulsation generating reciprocal active forces, melts solid epithelial tissues into a hyperuniform fluid state, while pulsating synchronization induces a reverse solidification. We demonstrate that the pulsation-induced hyperuniform fluid is a strict hyperuniform state that can recover from external perturbation. Our results also indicate that the reverse solidification is a BKT-type transition and the birth of topological defects in the pulsating phase space induces strong dynamical and structural heterogeneities at the transition point. We further construct a fluctuation hydrodynamic theory which can describe two opposite anomalous density fluctuations in the same framework.  Our work deepens our understanding of recently observed giant density fluctuation and topological defects in the MDCK monolayer~\cite{zehnder2015cell, hino2020erk, boocock2021theory, boocock2023interplay}. Although we have not yet found a biological tissue satisfying the condition for fluidic hyperuniformity, we expect that genetic engineering techniques may help to create synthetic cell tissue to realize this goal~\cite{lienert2014synthetic}. Moreover, the macroscopic pulsating robotic tissue~\cite{li2019particle} can also be used to test our prediction.

In addition, it would be straightforward to incorporate the mechanochemical feedback of cells in our model, which is expected to change the fluidization of tissue and corresponding BKT scenario and may induce other interesting fluctuation phenomena~\cite{zhang2023pulsating, boocock2023interplay, arzash2023tuning, banerjee2024hydrodynamics}. Moreover, it is also tempting to study pulsating cellular systems with nematic order and the relationship between nematic topological defects and tissue flow~\cite{saw2017topological,zhang2023active,rozman2023shape,chiang2024intercellular}. Such exploration would help us to understand how squeezing and contractile forces sculpt cell tissues~\cite{heisenberg2013forces, dance2021secret,bailles2022mechanochemical} and also guide the design of tissue-like robotic swarms~\cite{li2019particle}.

\section*{Materials and Methods}
\subsection*{Simulation details}
All simulations are performed within a square region $L^2$ with periodic boundaries. Due to the confluent nature of epithelial tissue, we have $L = \sqrt{NA_0}$ with unit area $A_0$. The center of cell $i$ obeys the overdamped dynamics, i.e. $ \dot{\mathbf{r}_i}  = \mu \mathbf{F}_i$ with $\mu$ the mobility coefficient and $\mathbf{F}_i = -\nabla_iE$ the intercellular force acting on cell $i$, where $E$ is the elastic energy of cell tissue. In our simulation, the unit length, unit energy and unit time are defined as $\sigma=\sqrt{A_0}$, $\epsilon_0=K_A A_0^2$ and $\tau= \sigma^2/\mu\epsilon_0$, respectively. Then, the dimensionless energy of the Voronoi model can be written as~\cite{fletcher2014vertex,bi2016motility,alt2017vertex,li2019mechanical}
\begin{eqnarray}
\epsilon &=& \frac{E}{\epsilon_0} \nonumber  \\
&=& \frac{1}{K_A A_0^2}\sum_i^N[K_A(A_i-A_0^i)^2 + K_P(P_i-P_0^i)^2] \nonumber \\
&=& \sum_i^N[(\frac{A_i}{A_0}-\frac{A_0^i}{A_0})^2 + \frac{K_P}{K_AA_0}(\frac{P_i}{\sqrt{A_0}}-\frac{P_0^i}{\sqrt{A_0}})^2] \nonumber \\
&=& \sum_i^N[(a_i-a_0^i)^2 + \xi(p_i-p_0^i)^2]. \label{energy}
\end{eqnarray}
Here, $K_A$ and $K_P$ are the area and perimeter modulus of cells. $A_i$ and $P_i$ are the area and perimeter of cell $i$, while  the corresponding $A^i_0$ and  $P^i_0$ are their preferred values. In this energy function, the first term on the right describes the area elasticity of the cell and the second originates from the competition between cell contractility and adhesion tension~\cite{bi2016motility,alert2020physical}. $\xi = K_P/(K_A A_0)$ is the dimensionless perimeter modulus. For area pulsation systems, we have $a_0^i=1 + \alpha \sin\varphi_i$ and $p_0^i=p_0$. For tension pulsation systems, we have $a_0^i=1$ and $p_0^i=p_0(1 + \beta \sin\varphi_i)$. 

Our simulation starts with a random initial configuration unless otherwise stated. We use Euler's method to integrate the equation of phase Eq.~(\ref{Kuramoto model}) and the equation of motion with a time step $\Delta t= 0.02\tau$. We fix $\xi=1$,  $T_\varphi=0.1 \tau^{-1}$ and $T_0=200 \tau$ in simulation unless otherwise stated. We simulate at least $10,000\tau$ to ensure that the system reaches a dynamically stable state before sampling. The simulation is performed with the number of cells $N$ ranging from $400$ to $262,144$ depending on the measured quantities. The Voronoi tesselation is based on the C++ library CGAL~\cite{fabri2009cgal}. 

\subsection*{Cage relative mean square displacement}
The dynamic state of cellular monolayers can be accurately characterized by the cage relative mean square displacement of cells, ${ {\rm MSD}_{c} }(t) = \frac{1}{N} \left\langle \sum_i |\mathbf{u}_i(t)- \mathbf{u}^{cage}_i(t)| \right\rangle$, which excludes the disturbance of long-wavelength fluctuations in 2D. Here, $\mathbf{u}_i(t)=\mathbf{r}_i(t)-\mathbf{r}_i(0)$ is the displacement of the cell $i$ over the time period $t$, and $\mathbf{u}^{cage}_i(t)=(1/n_i)\sum_{j \in C_i} \mathbf{u}_j(t)$ is the displacement of the `cage' which is composed of $n_i$ adjacent cells. $\left\langle \right\rangle$ represents the time average. The diffusion coefficient can be calculated by $D_c=\lim_{t \rightarrow \infty} [{ {\rm MSD}_{c} }(t)/(4t)]$.

\subsection*{Static structure factor}
The static structure factor is defined as $S(\mathbf{q}) = \frac{1}{N}\left\langle \left| \sum_j^N e^{i \mathbf{q} \cdot \mathbf{r}_j} \right|^2 \right\rangle$. Here, $\mathbf{q} = [q_x,q_y] = [i\frac{2\pi}{L},j\frac{2\pi}{L}]\ (i,j=1,2,3,...)$ and $L$ is the box size. Then, we project the vector $\mathbf{q}$ to scalar $q=\left| \mathbf{q} \right|$ and obtain the results of $S(q)$.

\subsection*{Other information}
The methods for calculation of crystal order, identification of topological defects, and preparation of single pair topological defects, as well as details of the hydrodynamic theory can be found in \emph{SI Appendix}.

\begin{acknowledgments} 
\subparagraph{Acknowledgments:} The authors are grateful to Ran Ni, Xia-qing Shi and Yan-Wei Li for their careful reading of the manuscript and helpful discussion.  This work is supported by  the National Natural Science Foundation of China (No. 12347102, 12275127, 12174184), the National Key Research and Development Program of China (No.~2022YFA1405000), the Innovation Program for Quantum Science and Technology (No.~2024ZD0300101), the Natural Science Foundation of Jiangsu Province (No. BK20233001). The simulations are performed on the High-Performance Computing Center of Collaborative Innovation Center of Advanced Microstructures, the High-Performance Computing Center (HPCC) of Nanjing University.
\end{acknowledgments} 

\bibliographystyle{nature}
\bibliography{PNASmain}

\begin{thebibliography}{88}
\expandafter\ifx\csname natexlab\endcsname\relax\def\natexlab#1{#1}\fi
\expandafter\ifx\csname url\endcsname\relax
  \def\url#1{\texttt{#1}}\fi
\expandafter\ifx\csname urlprefix\endcsname\relax\def\urlprefix{URL }\fi

\bibitem[{Friedl \& Gilmour(2009)}]{friedl2009collective}
Friedl, P. \& Gilmour, D.
\newblock Collective cell migration in morphogenesis, regeneration and cancer.
\newblock \emph{Nature Reviews Molecular Cell Biology} \textbf{10}, 445--457
  (2009).

\bibitem[{Hakim \& Silberzan(2017)}]{hakim2017collective}
Hakim, V. \& Silberzan, P.
\newblock Collective cell migration: a physics perspective.
\newblock \emph{Reports on Progress in Physics} \textbf{80}, 076601 (2017).

\bibitem[{Alert \& Trepat(2020)}]{alert2020physical}
Alert, R. \& Trepat, X.
\newblock Physical models of collective cell migration.
\newblock \emph{Annual Review of Condensed Matter Physics} \textbf{11}, 77--101
  (2020).

\bibitem[{Park \emph{et~al.}(2015)}]{park2015unjamming}
Park, J.-A. \emph{et~al.}
\newblock Unjamming and cell shape in the asthmatic airway epithelium.
\newblock \emph{Nature Materials} \textbf{14}, 1040--1048 (2015).

\bibitem[{Bi \emph{et~al.}(2015)Bi, Lopez, Schwarz \& Manning}]{bi2015density}
Bi, D., Lopez, J., Schwarz, J.~M. \& Manning, M.~L.
\newblock A density-independent rigidity transition in biological tissues.
\newblock \emph{Nature Physics} \textbf{11}, 1074--1079 (2015).

\bibitem[{Bi \emph{et~al.}(2016)Bi, Yang, Marchetti \&
  Manning}]{bi2016motility}
Bi, D., Yang, X., Marchetti, M.~C. \& Manning, M.~L.
\newblock Motility-driven glass and jamming transitions in biological tissues.
\newblock \emph{Physical Review X} \textbf{6}, 021011 (2016).

\bibitem[{Mongera \emph{et~al.}(2018)}]{mongera2018fluid}
Mongera, A. \emph{et~al.}
\newblock A fluid-to-solid jamming transition underlies vertebrate body axis
  elongation.
\newblock \emph{Nature} \textbf{561}, 401--405 (2018).

\bibitem[{Lin \emph{et~al.}(2018)Lin, Ye, Xu, Li \& Feng}]{lin2018dynamic}
Lin, S.-Z., Ye, S., Xu, G.-K., Li, B. \& Feng, X.-Q.
\newblock Dynamic migration modes of collective cells.
\newblock \emph{Biophysical Journal} \textbf{115}, 1826--1835 (2018).

\bibitem[{Giavazzi \emph{et~al.}(2018)}]{giavazzi2018flocking}
Giavazzi, F. \emph{et~al.}
\newblock Flocking transitions in confluent tissues.
\newblock \emph{Soft Matter} \textbf{14}, 3471--3477 (2018).

\bibitem[{Petrolli \emph{et~al.}(2019)}]{petrolli2019confinement}
Petrolli, V. \emph{et~al.}
\newblock Confinement-induced transition between wavelike collective cell
  migration modes.
\newblock \emph{Physical Review Letters} \textbf{122}, 168101 (2019).

\bibitem[{Peyret \emph{et~al.}(2019)}]{peyret2019sustained}
Peyret, G. \emph{et~al.}
\newblock Sustained oscillations of epithelial cell sheets.
\newblock \emph{Biophysical Journal} \textbf{117}, 464--478 (2019).

\bibitem[{Yu \emph{et~al.}(2021)}]{yu2021spatiotemporal}
Yu, J. \emph{et~al.}
\newblock Spatiotemporal oscillation in confined epithelial motion upon
  fluid-to-solid transition.
\newblock \emph{ACS Nano} \textbf{15}, 7618--7627 (2021).

\bibitem[{Barton \emph{et~al.}(2017)Barton, Henkes, Weijer \&
  Sknepnek}]{barton2017active}
Barton, D.~L., Henkes, S., Weijer, C.~J. \& Sknepnek, R.
\newblock Active vertex model for cell-resolution description of epithelial
  tissue mechanics.
\newblock \emph{PLOS Computational Biology} \textbf{13}, e1005569 (2017).

\bibitem[{Pasupalak \emph{et~al.}(2020)Pasupalak, Yan-Wei, Ni \&
  Ciamarra}]{pasupalak2020hexatic}
Pasupalak, A., Yan-Wei, L., Ni, R. \& Ciamarra, M.~P.
\newblock Hexatic phase in a model of active biological tissues.
\newblock \emph{Soft Matter} \textbf{16}, 3914--3920 (2020).

\bibitem[{Cai \emph{et~al.}(2022)Cai, Ji, Luo, Lei \& Ma}]{cai2022numerical}
Cai, L.-b., Ji, W., Luo, J., Lei, Q.-l. \& Ma, Y.-q.
\newblock Numerical study of dynamic zigzag patterns in migrating epithelial
  tissue.
\newblock \emph{Science China Physics, Mechanics \& Astronomy} \textbf{65},
  217011 (2022).

\bibitem[{Amiri \emph{et~al.}(2023)Amiri, Duclut, J{\"u}licher \&
  Popovi{\'c}}]{amiri2023random}
Amiri, A., Duclut, C., J{\"u}licher, F. \& Popovi{\'c}, M.
\newblock Random traction yielding transition in epithelial tissues.
\newblock \emph{Physical Review Letters} \textbf{131}, 188401 (2023).

\bibitem[{Tjhung \& Berthier(2017)}]{tjhung2017discontinuous}
Tjhung, E. \& Berthier, L.
\newblock Discontinuous fluidization transition in time-correlated assemblies
  of actively deforming particles.
\newblock \emph{Physical Review E} \textbf{96}, 050601 (2017).

\bibitem[{Zhang \& Fodor(2023)}]{zhang2023pulsating}
Zhang, Y. \& Fodor, {\'E}.
\newblock Pulsating Active Matter.
\newblock \emph{Physical Review Letters} \textbf{131}, 238302 (2023).

\bibitem[{Manacorda \& Fodor(2023)}]{manacorda2023pulsating}
Manacorda, A. \& Fodor, {\'E}.
\newblock Pulsating with discrete symmetry.
\newblock \emph{arXiv preprint arXiv:2310.14370}  (2023).

\bibitem[{Manning(2023)}]{manning2023essay}
Manning, M.~L.
\newblock Essay: Collections of deformable particles present exciting
  challenges for soft matter and biological physics.
\newblock \emph{Physical Review Letters} \textbf{130}, 130002 (2023).

\bibitem[{Liu \emph{et~al.}(2024{\natexlab{a}})Liu, Zhu \&
  Ai}]{liu2024collective}
Liu, W.-h., Zhu, W.-j. \& Ai, B.-q.
\newblock Collective motion of pulsating active particles in confined
  structures.
\newblock \emph{New Journal of Physics} \textbf{26}, 023017
  (2024{\natexlab{a}}).

\bibitem[{Pi{\~n}eros \& Fodor(2024)}]{pineros2024biased}
Pi{\~n}eros, W.~D. \& Fodor, {\'E}.
\newblock Biased ensembles of pulsating active matter.
\newblock \emph{arXiv preprint arXiv:2403.16961}  (2024).

\bibitem[{Martin \emph{et~al.}(2009)Martin, Kaschube \&
  Wieschaus}]{martin2009pulsed}
Martin, A.~C., Kaschube, M. \& Wieschaus, E.~F.
\newblock Pulsed contractions of an actin--myosin network drive apical
  constriction.
\newblock \emph{Nature} \textbf{457}, 495--499 (2009).

\bibitem[{Dierkes \emph{et~al.}(2014)Dierkes, Sumi, Solon \&
  Salbreux}]{dierkes2014spontaneous}
Dierkes, K., Sumi, A., Solon, J. \& Salbreux, G.
\newblock Spontaneous oscillations of elastic contractile materials with
  turnover.
\newblock \emph{Physical Review Letters} \textbf{113}, 148102 (2014).

\bibitem[{Lin \emph{et~al.}(2017)Lin, Li, Lan \& Feng}]{lin2017activation}
Lin, S.-Z., Li, B., Lan, G. \& Feng, X.-Q.
\newblock Activation and synchronization of the oscillatory morphodynamics in
  multicellular monolayer.
\newblock \emph{Proceedings of the National Academy of Sciences} \textbf{114},
  8157--8162 (2017).

\bibitem[{Curran \emph{et~al.}(2017)}]{curran2017myosin}
Curran, S. \emph{et~al.}
\newblock Myosin II controls junction fluctuations to guide epithelial tissue
  ordering.
\newblock \emph{Developmental Cell} \textbf{43}, 480--492 (2017).

\bibitem[{Krajnc \emph{et~al.}(2021)Krajnc, Stern \& Zankoc}]{krajnc2021active}
Krajnc, M., Stern, T. \& Zankoc, C.
\newblock Active instability and nonlinear dynamics of cell-cell junctions.
\newblock \emph{Physical Review Letters} \textbf{127}, 198103 (2021).

\bibitem[{P{\'e}rez-Verdugo \emph{et~al.}(2024)P{\'e}rez-Verdugo, Banks \&
  Banerjee}]{perez2024excitable}
P{\'e}rez-Verdugo, F., Banks, S. \& Banerjee, S.
\newblock Excitable dynamics driven by mechanical feedback in biological
  tissues.
\newblock \emph{Communications Physics} \textbf{7}, 167 (2024).

\bibitem[{Solon \emph{et~al.}(2009)Solon, Kaya-Copur, Colombelli \&
  Brunner}]{solon2009pulsed}
Solon, J., Kaya-Copur, A., Colombelli, J. \& Brunner, D.
\newblock Pulsed forces timed by a ratchet-like mechanism drive directed tissue
  movement during dorsal closure.
\newblock \emph{Cell} \textbf{137}, 1331--1342 (2009).

\bibitem[{Banerjee \emph{et~al.}(2015)Banerjee, Utuje \&
  Marchetti}]{banerjee2015propagating}
Banerjee, S., Utuje, K.~J. \& Marchetti, M.~C.
\newblock Propagating stress waves during epithelial expansion.
\newblock \emph{Physical Review Letters} \textbf{114}, 228101 (2015).

\bibitem[{Armon \emph{et~al.}(2018)Armon, Bull, Aranda-Diaz \&
  Prakash}]{armon2018ultrafast}
Armon, S., Bull, M.~S., Aranda-Diaz, A. \& Prakash, M.
\newblock Ultrafast epithelial contractions provide insights into contraction
  speed limits and tissue integrity.
\newblock \emph{Proceedings of the National Academy of Sciences} \textbf{115},
  E10333--E10341 (2018).

\bibitem[{Krajnc(2020)}]{krajnc2020solid}
Krajnc, M.
\newblock Solid--fluid transition and cell sorting in epithelia with junctional
  tension fluctuations.
\newblock \emph{Soft Matter} \textbf{16}, 3209--3215 (2020).

\bibitem[{Hino \emph{et~al.}(2020)}]{hino2020erk}
Hino, N. \emph{et~al.}
\newblock ERK-mediated mechanochemical waves direct collective cell
  polarization.
\newblock \emph{Developmental Cell} \textbf{53}, 646--660 (2020).

\bibitem[{Boocock \emph{et~al.}(2021)Boocock, Hino, Ruzickova, Hirashima \&
  Hannezo}]{boocock2021theory}
Boocock, D., Hino, N., Ruzickova, N., Hirashima, T. \& Hannezo, E.
\newblock Theory of mechanochemical patterning and optimal migration in cell
  monolayers.
\newblock \emph{Nature Physics} \textbf{17}, 267--274 (2021).

\bibitem[{Devany \emph{et~al.}(2021)Devany, Sussman, Yamamoto, Manning \&
  Gardel}]{devany2021cell}
Devany, J., Sussman, D.~M., Yamamoto, T., Manning, M.~L. \& Gardel, M.~L.
\newblock Cell cycle--dependent active stress drives epithelia remodeling.
\newblock \emph{Proceedings of the National Academy of Sciences} \textbf{118},
  e1917853118 (2021).

\bibitem[{Boocock \emph{et~al.}(2023)Boocock, Hirashima \&
  Hannezo}]{boocock2023interplay}
Boocock, D., Hirashima, T. \& Hannezo, E.
\newblock Interplay between mechanochemical patterning and glassy dynamics in
  cellular monolayers.
\newblock \emph{PRX Life} \textbf{1}, 013001 (2023).

\bibitem[{Yamamoto \emph{et~al.}(2022)Yamamoto, Sussman, Shibata \&
  Manning}]{yamamoto2022non}
Yamamoto, T., Sussman, D.~M., Shibata, T. \& Manning, M.~L.
\newblock Non-monotonic fluidization generated by fluctuating edge tensions in
  confluent tissues.
\newblock \emph{Soft Matter} \textbf{18}, 2168--2175 (2022).

\bibitem[{Duclut \emph{et~al.}(2022)Duclut, Paijmans, Inamdar, Modes \&
  J{\"u}licher}]{duclut2022active}
Duclut, C., Paijmans, J., Inamdar, M.~M., Modes, C.~D. \& J{\"u}licher, F.
\newblock Active T1 transitions in cellular networks.
\newblock \emph{The European Physical Journal E} \textbf{45}, 29 (2022).

\bibitem[{Staddon \emph{et~al.}(2022)Staddon, Munro \&
  Banerjee}]{staddon2022pulsatile}
Staddon, M.~F., Munro, E.~M. \& Banerjee, S.
\newblock Pulsatile contractions and pattern formation in excitable actomyosin
  cortex.
\newblock \emph{PLOS Computational Biology} \textbf{18}, e1009981 (2022).

\bibitem[{Thiagarajan \emph{et~al.}(2022)Thiagarajan, Bhat, Salbreux, Inamdar
  \& Riveline}]{thiagarajan2022pulsations}
Thiagarajan, R., Bhat, A., Salbreux, G., Inamdar, M.~M. \& Riveline, D.
\newblock Pulsations and flows in tissues as two collective dynamics with
  simple cellular rules.
\newblock \emph{iScience} \textbf{25} (2022).

\bibitem[{P{\'e}rez-Verdugo \& Banerjee(2023)}]{perez2023tension}
P{\'e}rez-Verdugo, F. \& Banerjee, S.
\newblock Tension Remodeling Regulates Topological Transitions in Epithelial
  Tissues.
\newblock \emph{PRX Life} \textbf{1}, 023006 (2023).

\bibitem[{Yin \emph{et~al.}(2024)Yin, Liu, Zhang, Liang \&
  Xu}]{yin2024emergence}
Yin, X., Liu, Y.-Q., Zhang, L.-Y., Liang, D. \& Xu, G.-K.
\newblock Emergence, Pattern, and Frequency of Spontaneous Waves in Spreading
  Epithelial Monolayers.
\newblock \emph{Nano Letters} \textbf{24}, 3631--3637 (2024).

\bibitem[{Liu \emph{et~al.}(2024{\natexlab{b}})Liu, Li, Wang \&
  Wu}]{liu2024emergence}
Liu, S., Li, Y., Wang, Y. \& Wu, Y.
\newblock Emergence of large-scale mechanical spiral waves in bacterial living
  matter.
\newblock \emph{Nature Physics} \textbf{20}, 1015--1021 (2024{\natexlab{b}}).

\bibitem[{Zehnder \emph{et~al.}(2015{\natexlab{a}})Zehnder, Suaris, Bellaire \&
  Angelini}]{zehnder2015cell}
Zehnder, S.~M., Suaris, M., Bellaire, M.~M. \& Angelini, T.~E.
\newblock Cell volume fluctuations in MDCK monolayers.
\newblock \emph{Biophysical Journal} \textbf{108}, 247--250
  (2015{\natexlab{a}}).

\bibitem[{Zehnder \emph{et~al.}(2015{\natexlab{b}})}]{zehnder2015multicellular}
Zehnder, S.~M. \emph{et~al.}
\newblock Multicellular density fluctuations in epithelial monolayers.
\newblock \emph{Physical Review E} \textbf{92}, 032729 (2015{\natexlab{b}}).

\bibitem[{Bocanegra-Moreno \emph{et~al.}(2023)Bocanegra-Moreno, Singh, Hannezo,
  Zagorski \& Kicheva}]{bocanegra2023cell}
Bocanegra-Moreno, L., Singh, A., Hannezo, E., Zagorski, M. \& Kicheva, A.
\newblock Cell cycle dynamics control fluidity of the developing mouse
  neuroepithelium.
\newblock \emph{Nature Physics} 1--9 (2023).

\bibitem[{Giavazzi \emph{et~al.}(2017)}]{giavazzi2017giant}
Giavazzi, F. \emph{et~al.}
\newblock Giant fluctuations and structural effects in a flocking epithelium.
\newblock \emph{Journal of Physics D: Applied Physics} \textbf{50}, 384003
  (2017).

\bibitem[{Jiao \emph{et~al.}(2014)}]{jiao2014avian}
Jiao, Y. \emph{et~al.}
\newblock Avian photoreceptor patterns represent a disordered hyperuniform
  solution to a multiscale packing problem.
\newblock \emph{Physical Review E} \textbf{89}, 022721 (2014).

\bibitem[{Chen \emph{et~al.}(2016)Chen, Aw, Devenport \&
  Torquato}]{chen2016structural}
Chen, D., Aw, W.~Y., Devenport, D. \& Torquato, S.
\newblock Structural characterization and statistical-mechanical model of
  epidermal patterns.
\newblock \emph{Biophysical Journal} \textbf{111}, 2534--2545 (2016).

\bibitem[{Li \emph{et~al.}(2018)Li, Das \& Bi}]{li2018biological}
Li, X., Das, A. \& Bi, D.
\newblock Biological tissue-inspired tunable photonic fluid.
\newblock \emph{Proceedings of the National Academy of Sciences} \textbf{115},
  6650--6655 (2018).

\bibitem[{Zheng \emph{et~al.}(2020)Zheng, Li \&
  Ciamarra}]{zheng2020hyperuniformity}
Zheng, Y., Li, Y.-W. \& Ciamarra, M.~P.
\newblock Hyperuniformity and density fluctuations at a rigidity transition in
  a model of biological tissues.
\newblock \emph{Soft Matter} \textbf{16}, 5942--5950 (2020).

\bibitem[{Liu \emph{et~al.}(2024{\natexlab{c}})Liu, Chen, Tian, Xu \&
  Jiao}]{liu2024universal}
Liu, Y., Chen, D., Tian, J., Xu, W. \& Jiao, Y.
\newblock Universal Hyperuniform Organization in Looped Leaf Vein Networks.
\newblock \emph{Physical Review Letters} \textbf{133}, 028401
  (2024{\natexlab{c}}).

\bibitem[{Torquato \& Stillinger(2003)}]{torquato2003local}
Torquato, S. \& Stillinger, F.~H.
\newblock Local density fluctuations, hyperuniformity, and order metrics.
\newblock \emph{Physical Review E} \textbf{68}, 041113 (2003).

\bibitem[{Torquato(2018)}]{torquato2018hyperuniform}
Torquato, S.
\newblock Hyperuniform states of matter.
\newblock \emph{Physics Reports} \textbf{745}, 1--95 (2018).

\bibitem[{Lei \emph{et~al.}(2019)Lei, Ciamarra \& Ni}]{lei2019nonequilibrium}
Lei, Q.-L., Ciamarra, M.~P. \& Ni, R.
\newblock Nonequilibrium strongly hyperuniform fluids of circle active
  particles with large local density fluctuations.
\newblock \emph{Science Advances} \textbf{5}, eaau7423 (2019).

\bibitem[{Lei \& Ni(2019)}]{lei2019hydrodynamics}
Lei, Q.-L. \& Ni, R.
\newblock Hydrodynamics of random-organizing hyperuniform fluids.
\newblock \emph{Proceedings of the National Academy of Sciences} \textbf{116},
  22983--22989 (2019).

\bibitem[{Ramaswamy \emph{et~al.}(2003)Ramaswamy, Simha \&
  Toner}]{ramaswamy2003active}
Ramaswamy, S., Simha, R.~A. \& Toner, J.
\newblock Active nematics on a substrate: Giant number fluctuations and
  long-time tails.
\newblock \emph{Europhysics Letters} \textbf{62}, 196 (2003).

\bibitem[{Shi \& Ma(2013)}]{shi2013topological}
Shi, X.-q. \& Ma, Y.-q.
\newblock Topological structure dynamics revealing collective evolution in
  active nematics.
\newblock \emph{Nature Communications} \textbf{4}, 3013 (2013).

\bibitem[{Newby \emph{et~al.}(2024)Newby, Shi, Jiao, Albert \&
  Torquato}]{newby2024structural}
Newby, E., Shi, W., Jiao, Y., Albert, R. \& Torquato, S.
\newblock Structural Properties of Hyperuniform Networks.
\newblock \emph{arXiv preprint arXiv:2411.06273}  (2024).

\bibitem[{Kuramoto(1975)}]{kuramoto1975international}
Kuramoto, Y.
\newblock International symposium on mathematical problems in theoretical
  physics.
\newblock \emph{Lecture Notes in Physics} \textbf{30}, 420 (1975).

\bibitem[{Zheng \emph{et~al.}(1998)Zheng, Hu \& Hu}]{zheng1998phase}
Zheng, Z., Hu, G. \& Hu, B.
\newblock Phase slips and phase synchronization of coupled oscillators.
\newblock \emph{Physical Review Letters} \textbf{81}, 5318 (1998).

\bibitem[{Acebr{\'o}n \emph{et~al.}(2005)Acebr{\'o}n, Bonilla, Vicente, Ritort
  \& Spigler}]{acebron2005kuramoto}
Acebr{\'o}n, J.~A., Bonilla, L.~L., Vicente, C. J.~P., Ritort, F. \& Spigler,
  R.
\newblock The Kuramoto model: A simple paradigm for synchronization phenomena.
\newblock \emph{Reviews of Modern Physics} \textbf{77}, 137 (2005).

\bibitem[{Rouzaire \& Levis(2021)}]{rouzaire2021defect}
Rouzaire, Y. \& Levis, D.
\newblock Defect superdiffusion and unbinding in a 2D XY model of self-driven
  rotors.
\newblock \emph{Physical Review Letters} \textbf{127}, 088004 (2021).

\bibitem[{Banerjee \emph{et~al.}(2024)Banerjee, Desaleux, Ranft \&
  Fodor}]{banerjee2024hydrodynamics}
Banerjee, T., Desaleux, T., Ranft, J. \& Fodor, {\'E}.
\newblock Hydrodynamics of pulsating active liquids.
\newblock \emph{arXiv preprint arXiv:2407.19955}  (2024).

\bibitem[{Li \& Li(2024)}]{li2024relaxation}
Li, M.-Y. \& Li, Y.-W.
\newblock Relaxation dynamics in the self-propelled Voronoi model for
  epithelial monolayers.
\newblock \emph{Physical Review Research} \textbf{6}, 033209 (2024).

\bibitem[{Ansell \emph{et~al.}(2024)Ansell, Li \& Sussman}]{ansell2024tunable}
Ansell, H.~S., Li, C. \& Sussman, D.~M.
\newblock Tunable glassy dynamics in models of dense cellular tissue.
\newblock \emph{arXiv preprint arXiv:2409.00496}  (2024).

\bibitem[{Huang \emph{et~al.}(2021)Huang, Hu, Yang, Liu \&
  Zhang}]{huang2021circular}
Huang, M., Hu, W., Yang, S., Liu, Q.-X. \& Zhang, H.
\newblock Circular swimming motility and disordered hyperuniform state in an
  algae system.
\newblock \emph{Proceedings of the National Academy of Sciences} \textbf{118},
  e2100493118 (2021).

\bibitem[{Gabrielli \emph{et~al.}(2008)Gabrielli, Joyce \&
  Torquato}]{gabrielli2008tilings}
Gabrielli, A., Joyce, M. \& Torquato, S.
\newblock Tilings of space and superhomogeneous point processes.
\newblock \emph{Physical Review E} \textbf{77}, 031125 (2008).

\bibitem[{Tang \emph{et~al.}(2024)Tang, Li \& Bi}]{tang2024tunable}
Tang, Y., Li, X. \& Bi, D.
\newblock Tunable Hyperuniformity in Cellular Structures.
\newblock \emph{arXiv preprint arXiv:2408.08976}  (2024).

\bibitem[{Hexner \& Levine(2017)}]{hexner2017noise}
Hexner, D. \& Levine, D.
\newblock Noise, diffusion, and hyperuniformity.
\newblock \emph{Physical Review Letters} \textbf{118}, 020601 (2017).

\bibitem[{Kuroda \& Miyazaki(2023)}]{kuroda2023microscopic}
Kuroda, Y. \& Miyazaki, K.
\newblock Microscopic theory for hyperuniformity in two-dimensional chiral
  active fluid.
\newblock \emph{Journal of Statistical Mechanics: Theory and Experiment}
  \textbf{2023}, 103203 (2023).

\bibitem[{Ikeda(2023)}]{ikeda2023harmonic}
Ikeda, H.
\newblock Harmonic chain far from equilibrium: single-file diffusion,
  long-range order, and hyperuniformity.
\newblock \emph{arXiv preprint arXiv:2309.03155}  (2023).

\bibitem[{Ikeda \& Kuroda(2024)}]{ikeda2024continuous}
Ikeda, H. \& Kuroda, Y.
\newblock Continuous symmetry breaking of low-dimensional systems driven by
  inhomogeneous oscillatory driving forces.
\newblock \emph{Physical Review E} \textbf{110}, 024140 (2024).

\bibitem[{Keta \& Henkes(2024)}]{keta2024long}
Keta, Y.-E. \& Henkes, S.
\newblock Long-range order in two-dimensional systems with fluctuating active
  stresses.
\newblock \emph{arXiv preprint arXiv:2410.14840}  (2024).

\bibitem[{Lienert \emph{et~al.}(2014)Lienert, Lohmueller, Garg \&
  Silver}]{lienert2014synthetic}
Lienert, F., Lohmueller, J.~J., Garg, A. \& Silver, P.~A.
\newblock Synthetic biology in mammalian cells: next generation research tools
  and therapeutics.
\newblock \emph{Nature Reviews Molecular Cell Biology} \textbf{15}, 95--107
  (2014).

\bibitem[{Li \emph{et~al.}(2019{\natexlab{a}})}]{li2019particle}
Li, S. \emph{et~al.}
\newblock Particle robotics based on statistical mechanics of loosely coupled
  components.
\newblock \emph{Nature} \textbf{567}, 361--365 (2019{\natexlab{a}}).

\bibitem[{Arzash \emph{et~al.}(2023)Arzash, Tah, Liu \&
  Manning}]{arzash2023tuning}
Arzash, S., Tah, I., Liu, A.~J. \& Manning, M.~L.
\newblock Tuning for fluidity using fluctuations in biological tissue models.
\newblock \emph{arXiv preprint arXiv:2312.11683}  (2023).

\bibitem[{Saw \emph{et~al.}(2017)}]{saw2017topological}
Saw, T.~B. \emph{et~al.}
\newblock Topological defects in epithelia govern cell death and extrusion.
\newblock \emph{Nature} \textbf{544}, 212--216 (2017).

\bibitem[{Zhang \& Yeomans(2023)}]{zhang2023active}
Zhang, G. \& Yeomans, J.~M.
\newblock Active forces in confluent cell monolayers.
\newblock \emph{Physical Review Letters} \textbf{130}, 038202 (2023).

\bibitem[{Rozman \emph{et~al.}(2023)Rozman, Yeomans \&
  Sknepnek}]{rozman2023shape}
Rozman, J., Yeomans, J.~M. \& Sknepnek, R.
\newblock Shape-tension coupling produces nematic order in an epithelium vertex
  model.
\newblock \emph{Physical Review Letters} \textbf{131}, 228301 (2023).

\bibitem[{Chiang \emph{et~al.}(2024)Chiang, Hopkins, Loewe, Marchetti \&
  Marenduzzo}]{chiang2024intercellular}
Chiang, M., Hopkins, A., Loewe, B., Marchetti, M.~C. \& Marenduzzo, D.
\newblock Intercellular friction and motility drive orientational order in cell
  monolayers.
\newblock \emph{Proceedings of the National Academy of Sciences} \textbf{121},
  e2319310121 (2024).

\bibitem[{Heisenberg \& Bella{\"\i}che(2013)}]{heisenberg2013forces}
Heisenberg, C.-P. \& Bella{\"\i}che, Y.
\newblock Forces in tissue morphogenesis and patterning.
\newblock \emph{Cell} \textbf{153}, 948--962 (2013).

\bibitem[{Dance(2021)}]{dance2021secret}
Dance, A.
\newblock The secret forces that squeeze and pull life into shape.
\newblock \emph{Nature} \textbf{589}, 186--189 (2021).

\bibitem[{Bailles \emph{et~al.}(2022)Bailles, Gehrels \&
  Lecuit}]{bailles2022mechanochemical}
Bailles, A., Gehrels, E.~W. \& Lecuit, T.
\newblock Mechanochemical principles of spatial and temporal patterns in cells
  and tissues.
\newblock \emph{Annual Review of Cell and Developmental Biology} \textbf{38},
  321--347 (2022).

\bibitem[{Fletcher \emph{et~al.}(2014)Fletcher, Osterfield, Baker \&
  Shvartsman}]{fletcher2014vertex}
Fletcher, A.~G., Osterfield, M., Baker, R.~E. \& Shvartsman, S.~Y.
\newblock Vertex models of epithelial morphogenesis.
\newblock \emph{Biophysical Journal} \textbf{106}, 2291--2304 (2014).

\bibitem[{Alt \emph{et~al.}(2017)Alt, Ganguly \& Salbreux}]{alt2017vertex}
Alt, S., Ganguly, P. \& Salbreux, G.
\newblock Vertex models: from cell mechanics to tissue morphogenesis.
\newblock \emph{Philosophical Transactions of the Royal Society B: Biological
  Sciences} \textbf{372}, 20150520 (2017).

\bibitem[{Li \emph{et~al.}(2019{\natexlab{b}})Li, Das \& Bi}]{li2019mechanical}
Li, X., Das, A. \& Bi, D.
\newblock Mechanical heterogeneity in tissues promotes rigidity and controls
  cellular invasion.
\newblock \emph{Physical Review Letters} \textbf{123}, 058101
  (2019{\natexlab{b}}).

\bibitem[{Fabri \& Pion(2009)}]{fabri2009cgal}
Fabri, A. \& Pion, S.
\newblock {CGAL}: the Computational Geometry Algorithms Library.
\newblock In \emph{Proceedings of the 17th ACM SIGSPATIAL international
  conference on advances in geographic information systems}, 538--539 (2009).

\end{thebibliography}

\end{document}